\documentclass[prd,aps,preprintnumbers,showpacs,twocolumn,floatfix,nofootinbib,superscriptaddress]{revtex4-1}
\usepackage{mathrsfs}
\usepackage{amsfonts}
\usepackage{amsmath}
\usepackage{array}
\usepackage{verbatim}
\usepackage{epsfig}
\usepackage{graphicx}

\usepackage{color} 

\newcommand*{\FigPath}{./}%

\begin{document}

\preprint{JLAB-THY-14-1965}
\preprint{LA-UR-14-28032}

\title{Nucleon tensor charge from Collins azimuthal asymmetry measurements}

\author{Zhong-Bo Kang}
\email{zkang@lanl.gov}
\affiliation{Theoretical Division,
                   Los Alamos National Laboratory,
                   Los Alamos, New Mexico 87545, USA}

\author{Alexei Prokudin}
\email{prokudin@jlab.org}
\affiliation{Jefferson Lab,
                   12000 Jefferson Avenue,
                   Newport News, Virginia 23606, USA}

\author{Peng Sun}
\email{psun@lbl.gov}
\affiliation{Nuclear Science Division,
                   Lawrence Berkeley National Laboratory,
                   Berkeley, California 94720, USA}

\author{Feng Yuan}
\email{fyuan@lbl.gov}
\affiliation{Nuclear Science Division,
                   Lawrence Berkeley National Laboratory,
                   Berkeley, California 94720, USA}

\begin{abstract}
We investigate the nucleon tensor charge from current experiments
by a combined analysis of the Collins asymmetries in two hadron
production in $e^+e^-$ annihilations and semi-inclusive hadron
production in deep inelastic scattering
processes. The
transverse momentum dependent
evolution is taken into
account, for the first time, in the global fit of the Collins fragmentation
functions and the quark transversity distributions at the approximate next-to-leading
logarithmic order. We obtain the nucleon tensor charge contribution
from up and down quarks as  $\delta u=+0.30^{+0.12}_{-0.08}$ and
$\delta d=-0.20_{-0.11}^{+0.28}$ at 90\% confidence level for momentum fraction $0.0065 \le x_B \le 0.35$ and $Q^2=$ 10 GeV$^2$.
\end{abstract}


\date{\today}
 \maketitle

\section{Introduction} 
Nucleon tensor charge is one of the fundamental properties
of the proton and its determination is among the main goals of
existing and future experimental facilities~\cite{Ralston:1979ys,Jaffe:1991kp,
Barone:2001sp,Dudek:2012vr,Boer:2011fh,Accardi:2012qut}.
It also plays an important role in constraining the nuclear physics aspects for
probing new physics beyond the standard model,
and has been an active subject from lattice QCD calculations~\cite{Bhattacharya:2011qm,Green:2012ej}.
In terms of the partonic structure of the nucleon, the tensor charge
is constructed from the quark transversity distribution,
one of the three leading-twist quark distributions.
However, the experimental exploration of the quark
transversity distribution in high energy scattering is difficult
because of its odd chirality~\cite{Jaffe:1991kp}.

An important channel is to
measure the Collins azimuthal asymmetries in semi-inclusive hadron
production in deep inelastic scattering (SIDIS),  where the transversity distribution
is coupled to the chiral-odd  Collins fragmentation function (FF)~\cite{Collins:1992kk},
as well as back-to-back two hadron production in $e^+e^-$ annihilations where two Collins FFs are
coupled to each other~\cite{Boer:1997mf}.

There have been great experimental efforts from both deep inelastic scattering (DIS) and $e^+e^-$
facilities to explore the Collins asymmetries, including
HERMES~\cite{Airapetian:2004tw,Airapetian:2010ds}, COMPASS~\cite{Adolph:2012sn}
and JLab~\cite{Qian:2011py} in DIS experiments, and BELLE~\cite{Abe:2005zx,Seidl:2008xc} and {\em BABAR}~\cite{TheBABAR:2013yha}
at $e^+e^-$ colliders of B-factories.
Due to the universality of the Collins fragmentation
functions~\cite{Metz:2002iz}, we will be able to combine the
analysis of these two processes to constrain the quark
transversity distributions.

Earlier results of the phenomenological studies in Refs.~\cite{Anselmino:2007fs,Anselmino:2008jk,Anselmino:2013vqa}
have demonstrated the powerful reach of the Collins
asymmetry measurements in accessing the quark transversity
distributions and eventually the nucleon tensor charge.
In this paper, we go beyond the leading order  
framework of Refs.~\cite{Anselmino:2007fs,Anselmino:2008jk,Anselmino:2013vqa},
and take into account the important higher order corrections, including,
in particular, the large logarithms~\cite{Collins:1981uk,Collins:1984kg}.
Theoretically, the large logarithms in the above hard processes
are controlled by the relevant QCD evolution, i.e., the
transverse momentum dependent (TMD) evolution~\cite{Collins:1981uk,Collins:1984kg}.
It was pointed out in Ref.~\cite{Boer:2001he} that the TMD evolution
plays an important role in evaluating the Collins asymmetries.
Because of the large energy difference between the existing DIS and $e^+e^-$
experiments~\cite{Airapetian:2004tw,Airapetian:2010ds,Adolph:2012sn,Qian:2011py,Abe:2005zx,Seidl:2008xc,TheBABAR:2013yha},
the QCD evolution effects have to be carefully examined when one extracts
the quark transversity distributions.
In this paper, for the first time, we demonstrate that the TMD evolution can
describe the experimental data and constrain the nucleon tensor charge with
improved theoretical accuracy.
To achieve that, we include the most recent developments from both theory and phenomenology
sides~\cite{Collins:2011zzd,Yuan:2009dw,Kang:2010xv, Kang:2011mr,Echevarria:2012js,
Bacchetta:2013pqa,Sun:2013hua,Echevarria:2014xaa,Echevarria:2014rua,Su:2014wpa}
and apply the TMD evolution at the next-to-leading-logarithmic (NLL) order within
the Collins-Soper-Sterman (CSS)~\cite{Collins:1981uk,Collins:1984kg} formalism.
We show that our results improve the theoretical description of the experimental
data in various aspects, especially, in formulating the transverse momentum
dependence of the asymmetries in $e^+e^-$ annihilations~\cite{TheBABAR:2013yha}.
The quark transversity distribution has also been an important subject in exploring
other transverse spin related phenomena, such as the dihadron fragmentation
processes~\cite{Bacchetta:2012ty,Bacchetta:2011ip}, and inclusive hadron production
at large transverse momentum in single transversely polarized $pp$
collisions~\cite{Kang:2010zzb,Anselmino:2012rq,Kanazawa:2014dca}.
Our results will provide an important cross-check and a step further
toward a global analysis of all these spin asymmetries associated
with the quark transversity distributions.
\section{Collins Asymmetries in SIDIS and $e^+e^-$ annihilation} 
In SIDIS, a lepton scatters off the nucleon target $N$, and produces
an identified hadron $h$ in the final state,
$l N \to lh X  $.
The Collins effect leads to a transverse spin asymmetry: $\sigma(S_\perp) \sim F_{UU}(1+A_{UT}^{\sin(\phi_h+\phi_s)}\sin(\phi_h+\phi_s))$, where $\phi_{s}$  and $\phi_{h}$ are the azimuthal angles of the nucleon's
transverse polarization vector $\vec{S}_\perp$ and the transverse
momentum vector $\vec{P}_{h\perp}$ of the final-state hadron, respectively.
The asymmetry $A_{UT}^{\sin(\phi_h+\phi_s)}$ can be calculated as
\begin{equation}
A_{UT}^{\sin(\phi_h+\phi_s)}(x_B, y, z_h, P_{h\perp})  =   \frac{2 (1-y)}{1+(1-y)^2}
\frac{F_{UT}}{F_{UU}} \ ,
\end{equation}
with usual SIDIS kinematic variables $x_B$, $y$, $z_h$, and $Q^2\simeq x_B y\, S$, and $S$ is the lepton-nucleon center of mass energy.
The structure functions $F_{UU}$ ($F_{UT}$) depend on the kinematic
variables and can be factorized into the TMD quark distribution (transversity)
and fragmentation (Collins) functions in the low transverse momentum
region. Applying the TMD evolution, we can write down $F_{UU},F_{UT}$   as \cite{Collins:1981uk,Collins:1984kg,Boer:2001he,Kang:2011mr,tobe}
\begin{eqnarray}
F_{UU} &=& \frac{1}{z_h^2}\int \frac{db\, b}{2\pi} J_0\!\!\left( \frac{{P}_{h\perp} b}{z_h} \right)\,
e^{-S_{\rm PT}(Q,b_*)-S_{\rm NP}^{\rm (SIDIS)}(Q,b)}  \nonumber\\
&\times & \, C_{q\leftarrow i}\otimes f_{1}^{i}(x_B,\mu_b)  \,\,\,
\hat{C}_{j\leftarrow q}^{\rm (SIDIS)}\otimes \hat D_{h/j}(z_h,\mu_b)  \label{fuu},\\
F_{UT} &=& -\frac{1}{2 z_h^3} \int \frac{db \, b^2}{2\pi} J_1\!\!\left( \frac{{P}_{h\perp} b}{z_h} \right)\,
e^{-S_{\rm PT}(Q,b_*)-S_{\rm NP\, coll}^{\rm (SIDIS)}(Q,b)} \nonumber\\
&\times & \, \delta C_{q\leftarrow i}\otimes h_{1}^{i}(x_B,\mu_b)  \,\,\,
\delta \hat{C}_{j\leftarrow q}^{\rm (SIDIS)}\otimes \hat H_{h/j}^{(3)}(z_h,\mu_b) , \label{eq:fut1}
\end{eqnarray}
where $b$ is Fourier conjugate variable to the measured final hadron
momentum ${P}_{h\perp}$, $J_1$ is
the Bessel function, $\mu_b=c_0/b_*$ with $c_0\simeq 1.12$, and the symbol
$\otimes$ represents the usual convolution in momentum fractions.
Summation over quark flavors $q$ weighted with quark
charge $\sum_q e^2_q$ and summation over $i,j = q,\bar q, g$ is
implicit in all formulas for structure functions. {$C$,
$\hat{C} $ and $\delta C$, 
$\delta \hat{C} $ are coefficient functions for unpolarized distribution, fragmentation function, and transversity and Collins
FF that can be calculated perturbatively.

The $b_*$-prescription ($b$ $\to$ $b_*$ $\equiv$ $b/\sqrt{1+b^2/b_{max}^2}$ with $b_{max}$ =1.5 GeV$^{-1}$ in
our calculations) was applied to introduce the nonperturbative form factors 
$S_{\rm NP}^{\rm (SIDIS)}$ and $S_{\rm NP\, coll}^{\rm (SIDIS)}$ that contain information on initial conditions of evolution.
The Collins fragmentation function~\cite{Collins:1992kk} enters as the transverse momentum moment~\cite{Yuan:2009dw}, $\hat{H}_{h/q}^{(3)}(z_h)=\int d^2p_\perp \frac{|p_\perp^2|}{M_h} H_{1\, h/q}^\perp(z_h,p_\perp)$, 
where $H_{1\, h/q}^\perp(z_h,p_\perp)$ is the quark Collins function defined in \cite{Yuan:2009dw}, and differs by a factor of $\left(-1/{z_h}\right)$ from the so-called ``Trento convention''~\cite{Bacchetta:2004jz},
\begin{equation}
H_{1\, h/j}^\perp(z_h,p_\perp)= -\frac{1}{z_h}H_{1\, h/j}^\perp(z_h,p_\perp)|_{\rm Trento},
\end{equation}
with $p_\perp$ the transverse component of the hadron with respect to the fragmenting quark momentum.

Three important ingredients have to be included to achieve the NLL
formalism for the above structure functions and asymmetries.
First, the perturbative Sudakov form factor~\cite{Koike:2006fn},
\begin{equation}
S_{\rm PT}(Q,b_*)=\int_{\mu_b^2}^{Q^2}\frac{d\mu^2}{\mu^2}\left[A\ln\frac{Q^2}{\mu^2}+B\right] \ ,\label{eq:spert}
\end{equation}
with perturbative coefficients $A^{(1,2)}\sim{\alpha_s^{(1,2)}}$ and $B^{(1)}\sim{\alpha_s^1}$~\cite{Nadolsky:1999kb,Koike:2006fn}.
Second, the scale evolutions of the quark transversity distribution
and the Collins fragmentation functions up to the scale of $\mu_b$.
The evolution for the quark transversity is known
\begin{equation}
\frac{\partial}{\partial \ln\mu^2}h_{1}^q(x,\mu)=\frac{\alpha_s}{2\pi}{P}_{q\leftarrow q}^{h_1}
\otimes h_1^q(x,\mu)\ ,
\end{equation}
with the splitting kernel ${P}_{q\leftarrow q}^{h_1}$ given in~\cite{Artru:1989zv}.
The evolution equation for $\hat H^{(3)}_{h/q}$ is more
complicated~\cite{Yuan:2009dw,Kang:2010xv,Kanazawa:2013uia}. However, if we
keep only the homogenous term, it reduces to a simpler form as
\begin{align}
\frac{\partial}{\partial \ln\mu^2}\hat H^{(3)}_{h/q}(z_h,\mu)=\frac{\alpha_s}{2\pi}{P}_{q\leftarrow q}^{\rm coll}
\otimes \hat H^{(3)}_{h/q}(z_h,\mu)\ ,
\end{align}
and it is interesting to find out that the splitting kernel ${P}_{q\leftarrow q}^{\rm coll}$ for the
homogenous term is the same~\cite{Kang:2010xv} as that for
the quark transversity distribution. As a first study, we will use this approximation and call resulting resummation NLL$'$.

Third, the $C$-coefficients are calculated at one-loop order ($C^{(1)}$)~\cite{Nadolsky:1999kb,Koike:2006fn},
for which we have~\cite{Yuan:2009dw,Bacchetta:2013pqa,Echevarria:2014rua,tobe}:
$\delta {C}_{q\leftarrow q}^{(1)}(x, \mu_b)=\frac{\alpha_s}{\pi}\left(-2C_F\delta(1-x)\right)$
and $\delta \hat C_{q\leftarrow q}^{\rm (SIDIS)(1)} (z, \mu_b)=\frac{\alpha_s}{\pi}\left({P}_{q\leftarrow q}^{\rm coll}(z) \ln z -2C_F\delta(1-z)\right)$.
Again, we only keep the homogenous term in the latter coefficient.
In the CSS formalism, there is a freedom to include part of $C$-coefficient
contributions into a hard factor~\cite{Collins:2011zzd,Catani:2000vq},
and the difference is in higher next-to-next-to-leading-logarithmic order (NNLL). This difference is negligible in
our numeric calculations.

In the two hadron productions in $e^+e^-$ annihilations, $e^++e^-\to h_1+h_2+X$, a quark-antiquark pair is
produced and fragments into hadrons, where two of them are observed in the final state in opposite hemispheres.
The center of mass energy $S=Q^2=(P_{e^+}+P_{e^-})^2$, and the final-state
two hadrons have momenta $P_{h1}$ and $P_{h2}$, respectively. The Collins effect
leads to an azimuthal angular $\cos\left(2\phi_0\right)$ asymmetry between the two
hadrons~\cite{Boer:1997mf}, and can be quantified as
\begin{eqnarray}
R^{h_1h_2} \equiv
  1+\cos(2 \phi_0) \frac{ \sin^2\theta }{ 1+\cos^2\theta  }
  \frac{Z_{\rm coll}^{h_1h_2}}{Z_{uu}^{h_1h_2}} \ ,
\end{eqnarray}
where $\theta$ is the polar angle between the hadron $h_2$ and the beam direction of
$e^+e^-$, and $\phi_0$ is defined as the azimuthal
angle of hadron $h_1$ relative to that of hadron $h_2$. To cancel possible acceptance effects as well as radiative effects, experiments measure the so-called double ratio asymmetries $A_0$ and $A_{12}$
, which are related to the ratios of $R^{h_1 h_2}$ from different hadron pair combinations, for details, see \cite{Abe:2005zx,Seidl:2008xc,TheBABAR:2013yha}.
 In the current study, we focus on the so-called $A_0$~\cite{Abe:2005zx,Seidl:2008xc,TheBABAR:2013yha} asymmetry. With TMD evolution included, the final results for $Z$ functions are given by \cite{Boer:2001he,tobe},
\begin{align}
Z_{uu}^{h_1h_2} =&\frac{1}{z_{h1}^2}\int \frac{db\, b}{(2\pi)} J_0\!\!\left(  \frac{{P}_{h\perp} {b}}{z_{h1}}\right) \;
e^{-{S}_{\rm PT}(Q,b_*)-S_{\rm NP}^{(e^+e^-)}(Q,b)}
\nonumber\\
&\times
\hat{C}_{i\leftarrow q}^{(e^+e^-)}\otimes D_{h_1/i}(z_{h1},\mu_b)
\nonumber \\
&\times \hat{C}_{j\leftarrow \bar q}^{(e^+e^-)}\otimes D_{h_2/j}(z_{h2},\mu_b) \label{eq:zuu1} \ ,
\\
Z_{\rm coll}^{h_1h_2} = & \frac{1}{z_{h1}^2}\frac{1}{4z_{h1}z_{h2}}\int \frac{db\, b^3}{(2\pi)} J_2\!\!\left(  \frac{{P}_{h\perp} {b}}{z_{h1}}\right)  \;
e^{-{S}_{\rm PT}(Q,b_*)}
\nonumber\\
&\times e^{-S_{\rm NP\ coll}^{(e^+e^-)}(Q,b)} \,\,\, \delta \hat C_{i\leftarrow q}^{(e^+e^-)}\otimes \hat H^{(3)}_{h_1/i}(z_{h1},\mu_b)
\nonumber\\
&\times \delta \hat{C}_{j\leftarrow \bar q}^{(e^+e^-)}\otimes \hat H_{h_2/j}^{(3)}(z_{h2},\mu_b)
\label{eq:zut1} \ ,
\end{align}
where $z_{hi}=2|P_{hi}|/Q$, $P_{h\perp}$ is the transverse momentum of hadron $h_1$, and the coefficient for the Collins function at one-loop order is given by
$\delta \hat{C}_{q\leftarrow q}^{(e^+e^-)(1)}(z, \mu_b)=\frac{\alpha_s}{\pi}\left({P}_{q\leftarrow q}^{\rm coll}(z) \ln z +\frac{C_F}{4}\left(\pi^2 -8\right)\delta(1-z)\right)$, while the coefficients $\hat C_{j\leftarrow q}^{(e^+e^-)(1)}(z, \mu_b)$ are derived in~\cite{Collins:1985xx,tobe}.  The TMD factorization for the so-called $A_{12}$ asymmetry cannot be straightforwardly formulated  ~\cite{tobe} because of additional requirement
of jet axis involved in experiments.

\section{Global analysis with TMD evolution}  
To perform the global analysis of the experimental data, we should parametrize the nonperturbative form factors.
For the spin-averaged cross sections, we
follow the parametrizations in Ref.~\cite{Su:2014wpa},
\begin{eqnarray}
S_{\rm NP}^{\rm(SIDIS)}&=&g_2 \ln\left({b}/{b_*}\right)\ln\left({Q}/{Q_0}\right)+ \nonumber  \\
 & & \left({g_q} +  {g_h}/{z_h^2}\right)b^2 \ ,\\
S_{\rm NP}^{\rm(e^+e^-)}&=&g_2 \ln\left(b/b_*\right)\ln(Q/Q_0)+ \nonumber \\
& & g_hb^2\left(1/z_{h1}^2+1/z_{h2}^2\right) \ ,
\end{eqnarray}
where the initial scale is chosen to be  $Q_0^2$ = 2.4 GeV$^2$, and other parameters are  determined from
the analysis of unpolarized SIDIS and Drell-Yan processes in Ref.~\cite{Su:2014wpa}: $g_q = g_1/2=0.106$, $g_2=0.84$, $g_h=0.042$  (GeV$^2$).   The presence of $\alpha_s^1$ contributions to $C$-coefficients requires normalization factors in the fit of  Ref.~\cite{Su:2014wpa}, however they affect both polarized and unpolarized parts equally thus there is no need for any additional normalization factor in the asymmetry.
The parametrization of $S_{\rm NP}^{(e^+e^-)}$ follows the universality arguments of
the TMDs. For the Collins asymmetries, we need to take into account different initial conditions for transversity and Collins FF.
We introduce a new parameter, $g_c$, to take into account the different $b$-shape of the Collins fragmentation function and
write, using universality of the Collins function between these two processes,
\begin{eqnarray}
S_{\rm NP\ coll}^{\rm(SIDIS)}&=&S_{\rm NP}^{\rm (SIDIS)}-g_c b^2/z_h^2\ ,\\
S_{\rm NP\ coll}^{\rm(e^+e^-)}&=&S_{\rm NP}^{\rm (e^+e^-)}-g_cb^2\left(1/z_{h1}^2+1/z_{h2}^2\right) \ .
\end{eqnarray}
In the global fit, we parametrize the quark transversity distributions at the initial scale $Q_0$
to satisfy the Soffer bound~\cite{Soffer:1995ww,Vogelsang:1997ak} as,
\begin{align}
h_1^{q}(x,Q_0)=&N_{q}^h x^{a_{q}}(1-x)^{b_{q}} \frac{(a_{q} + b_{q})^{a_{q} + b_{q}}}
{a_{q}^{a_{q}} b_{q}^{b_{q}}}
\nonumber \\
&\times \frac{1}{2}\left (f_{1}(x,Q_0) + g_{1}(x,Q_0)  \right ) \ ,
\end{align}
 with $\left|N_q^h\right|\leq 1$ for up and down quarks $q=u,d$, respectively, where $f_{1}$ are the unpolarized
CT10 next-to-leading order (NLO) quark distributions~\cite{Lai:2010vv} and $g_{1}$ are the DSSV helicity NLO distributions~\cite{deFlorian:2009vb}.
In the current study, we assume all the sea quark transversity distributions
are negligible.

Similarly, we parametrize
the moments for the Collins fragmentation functions in terms of the unpolarized fragmentation functions,
\begin{align}
\hat{H}_{fav}^{(3)}(z,Q_0)&= N_{u}^c z^{\alpha_{u}}(1-z)^{\beta_{u}} D_{\pi^+/u}(z,Q_0) \ , \label{eq:fav}\\
\hat{H}_{unf}^{(3)}(z,Q_0)&= N_{d}^c z^{\alpha_{d}}(1-z)^{\beta_{d}} D_{\pi^+/d}(z,Q_0) \ ,  \label{eq:unfav}
\end{align}
for the favored and unfavored Collins
fragmentation functions, respectively.
The rest can be obtained by applying the isospin relations.
We also neglect possible difference of favored/unfavored fragmentation
functions of $\bar u, \bar d$ and $u, d$. In our fit, we include the strange quark Collins FF,
which is parametrized similar to unfavored function in Eq.~(\ref{eq:unfav})
with unpolarized strange FF. We also utilize the newest NLO extraction of fragmentation functions~\cite{deFlorian:2014xna}. The new DSS FF set
 is capable of describing pion multiplicities measured by the COMPASS and HEMRES collaborations.

\begin{table}[htb]
\begin{tabular}{l c l l c l l c l}
\hline
 & & & & & & & &\\
$N_u^h$ &=& $0.85\pm 0.09$ & $a_u$ &=& $ 0.69 \pm 0.04$ & $b_u$ &=& $ 0.05 \pm 0.04$ \\
$N_d^h$ &=& $-1.0\pm 0.13$ & $a_d$ &=& $ 1.79 \pm 0.32$ & $b_d$ &=& $ 7.00 \pm 2.65$  \\
$N_u^c$ &=& $-0.262\pm 0.025$ & $\alpha_u$ &=& $ 1.69 \pm 0.01$ & $\beta_u$ &=& $ 0.00 \pm 0.54$ \\
$N_d^c$ &=& $0.195\pm 0.007$ & $\alpha_d$ &=& $ 0.32 \pm 0.04$ & $\beta_d$ &=& $ 0.00 \pm 0.79$ \\
$g_c$ &=& $0.0236\pm 0.0007$&\multicolumn{3}{l}{(GeV$^2$)}\\
& & & & & & & &\\
\hline
& & & & & & & &\\
\multicolumn{3}{l}{$\chi_{min}^2 =  218.407$} & \multicolumn{3}{l}{$\chi^2_{min}/{n.d.o.f}=0.88$}  \\
\end{tabular}
\caption{Fitted parameters of the transversity quark distributions for $u$ and $d$ and Collins fragmentation functions.
The fit is performed by using the MINUIT minimization package. Quoted errors correspond to the MINUIT estimate.}
\label{parameters}
\end{table}

In total we have 13 parameters in our global fit:
$N_u^h$, $N_d^h$, $a_u$, $a_d$, $b_u$, $b_d$, $N_u^c$, $N_d^c$,
$\alpha_u$, $\alpha_d$, $\beta_u$, $\beta_d$, $g_c$.
In the fit, we include all existing SIDIS data ($n_{SIDIS}=140$ points), all points in $x_B$, $z_h$, and $P_{h\perp}$  where the formalism is valid (we limit $P_{h\perp} < 0.8$ GeV) for $\pi^\pm$ pion production from HERMES~\cite{Airapetian:2004tw,Airapetian:2010ds}, COMPASS~\cite{Adolph:2012sn} and JLab HALL A ~\cite{Qian:2011py}.
For the Collins asymmetries in $e^+e^-$ annihilation experiments we have $n_{e^+e^-}=122$ data points, measurements as function of $z_{h1}$, $z_{h2}$, and $P_{h\perp}$ (we limit $P_{h\perp}/z_{h1} < 3.5$ GeV) from BELLE~\cite{Seidl:2008xc}
and {\em BABAR}~\cite{TheBABAR:2013yha} collaborations. We use the MINUIT minimization package to perform the fit. The resulting parameters are presented in Table~\ref{parameters}.
The total $\chi^2 =  218.407 $, $n_{d.o.f.} = 249$, and $\chi^2/n_{d.o.f} = 0.88$. The fit is equally good for SIDIS and $e^+e^-$ data
 $\chi^2_{SIDIS}/{n_{SIDIS}} =  0.93$, $\chi^2_{e^+e^-}/{n_{e^+e^-}} =  0.72$.  The goodness of resulting fit is 90\%~\cite{tobe,Press:1992zz} and inclusion of more parameters
 does not improve it. We estimate flavor dependence of functions by allowing a flavor dependent functional form. Note that our resulting $d$ quark transversity is very close to its bound, the same feature was also found in Refs.~\cite{Bacchetta:2012ty,Bacchetta:2011ip}.
We plot the extracted transversity and Collins fragmentation function in Fig.~\ref{fig:functions} at two different scales $Q^2 = 10$ and 1000 GeV$^2$. Only relative sign of transversity can be determined and we present here a solution with positive $u$ quark transversity as in Refs.~\cite{Anselmino:2007fs,Anselmino:2008jk,Anselmino:2013vqa,Bacchetta:2012ty,Bacchetta:2011ip}. Favorite and unfavorite Collins FFs are of opposite signs as suggested by the sum rules \cite{Schafer:1999kn,Meissner:2010cc}.

\begin{figure}[tbp]
\centering
\includegraphics[width=4.2cm,bb= 100 10 510 500]{\FigPath/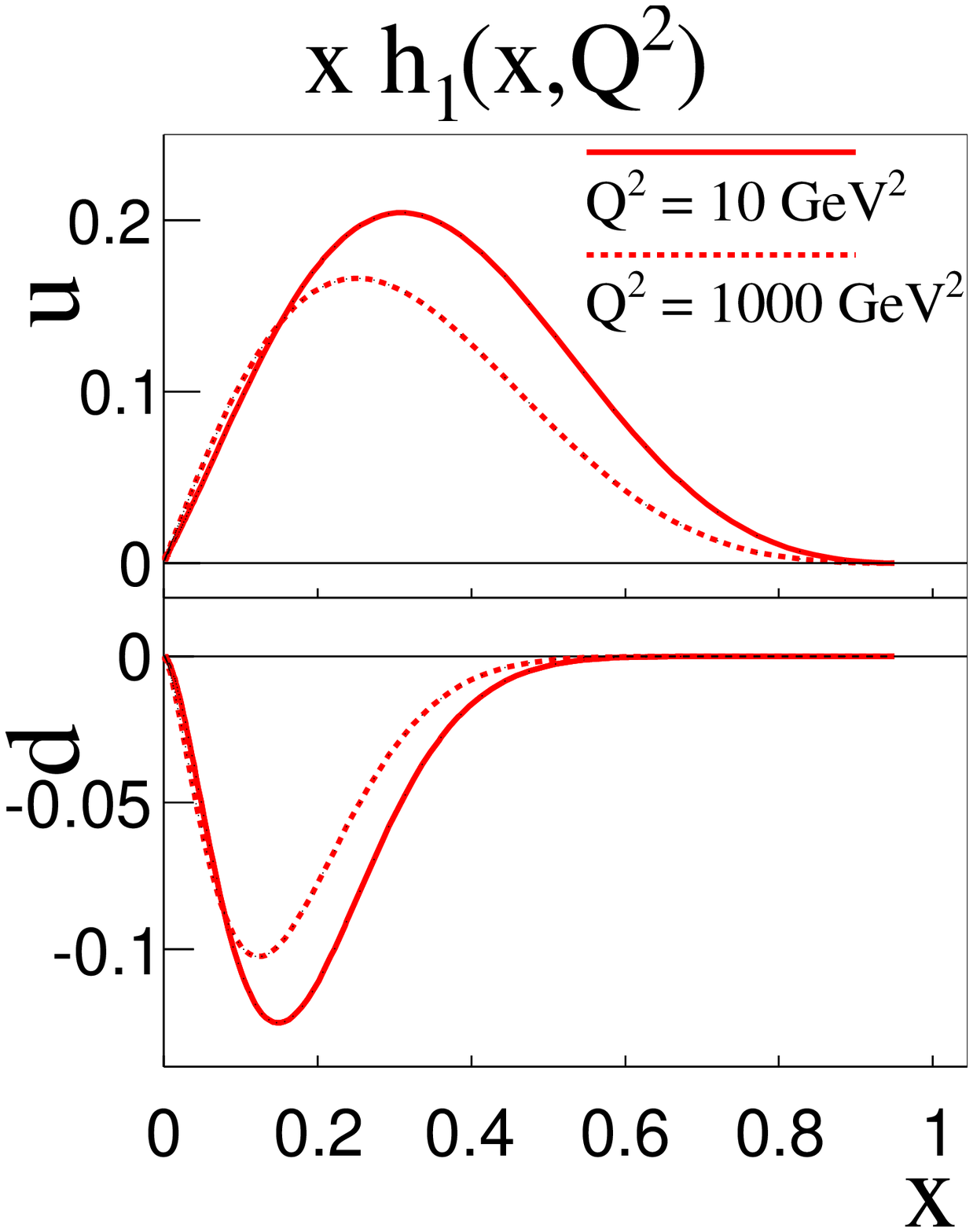}
\includegraphics[width=4.2cm,bb= 100 10 490 500]{\FigPath/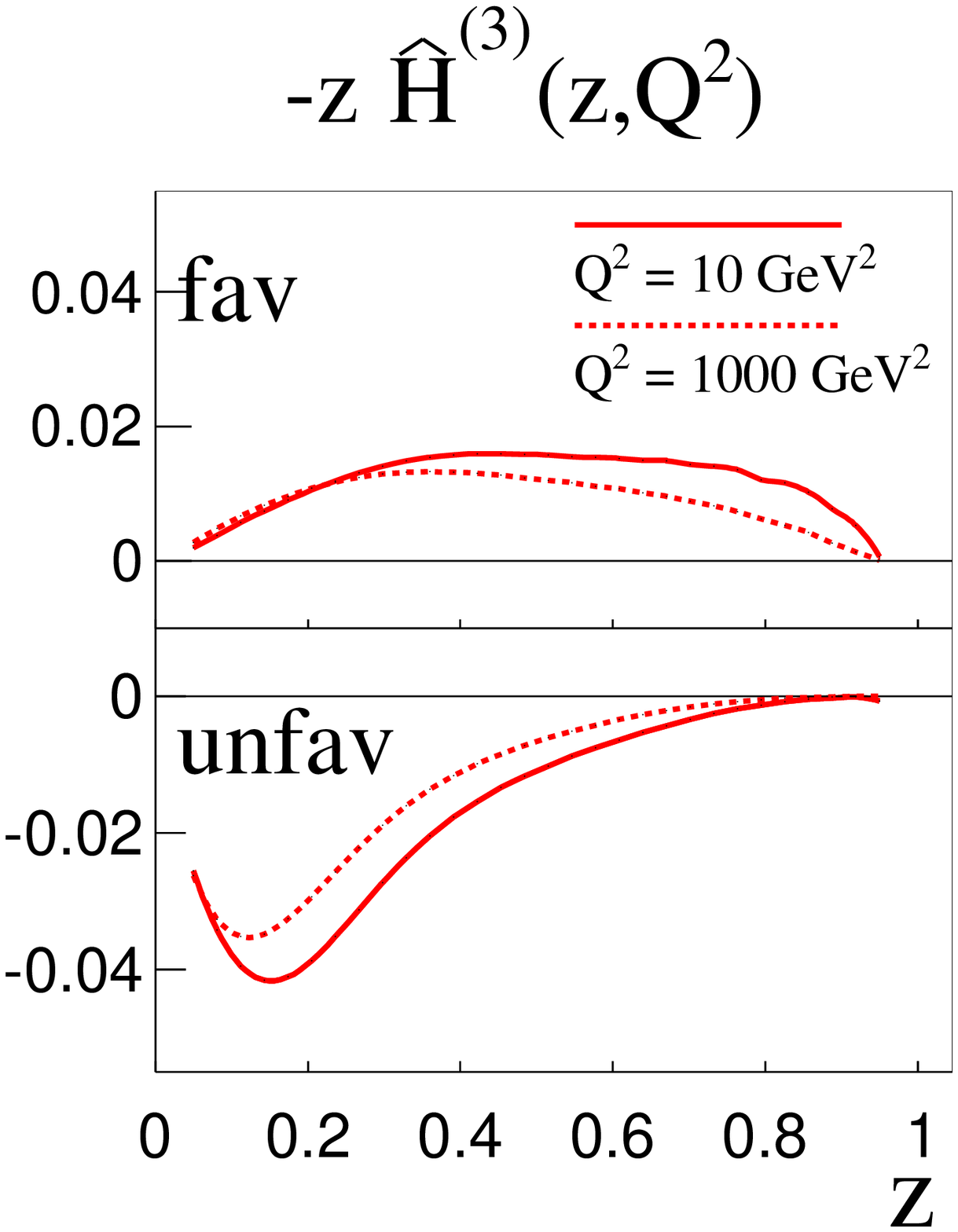}
\caption{Extracted transversity distribution  and Collins fragmentation function at two different scales  $Q^2 = 10$ (solid lines) and $Q^2 = 1000$ (dashed lines) GeV$^2$.}
\label{fig:functions}
\end{figure}

We also show an example of description of experimental data, namely $P_{h\perp}$ dependence of asymmetry in $e^+e^-$ from the {\em BABAR}~\cite{TheBABAR:2013yha} collaboration in Fig.~\ref{fig:babar}. One can see that NLL$'$ accuracy adequately describes the data.  In this plot we also show theoretical computations without TMD evolution (dotted line), leading-logarithmic  (LL) accuracy (dashed line),
and the complete NLL$'$ accuracy (solid line). The difference between these computations diminishes when we include higher orders, it means that the theoretical uncertainty
improves. We conjecture that the difference between NLL$'$ and NNLL will be smaller than difference between NLL$'$ and LL and thus be comparable to experimental errors. One can also observe that asymmetry at $Q^2=110$ GeV$^2$ is suppressed by a factor of 2 to 3  with respect to tree-level calculations due to the Sudakov form factor.
\begin{figure}[tbp]
\centering
\includegraphics[width=7cm]{\FigPath/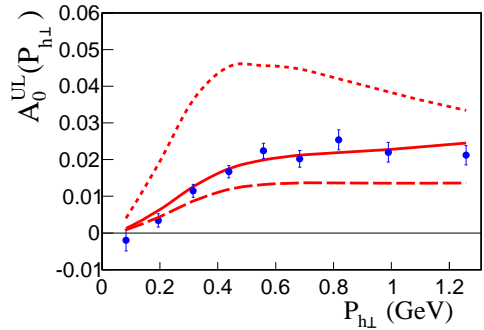}
\caption{Collins asymmetries measured by the {\em BABAR}~\cite{TheBABAR:2013yha} Collaboration as a function of
$P_{h\perp}$ in production of unlike sign ``U" over like sign ``L" pion pairs at $Q^2 = 110$ GeV$^2$.
The solid line corresponds to the full NLL$'$ calculation, the dashed line to the LL calculation, and the
dotted to the calculation without TMD evolution. Calculations are performed with parameters from Table~\ref{parameters}. }
\label{fig:babar}
\end{figure}

\begin{figure}[tbp]
\centering
\includegraphics[width=4.2cm,bb= 50 10 595 550]{\FigPath/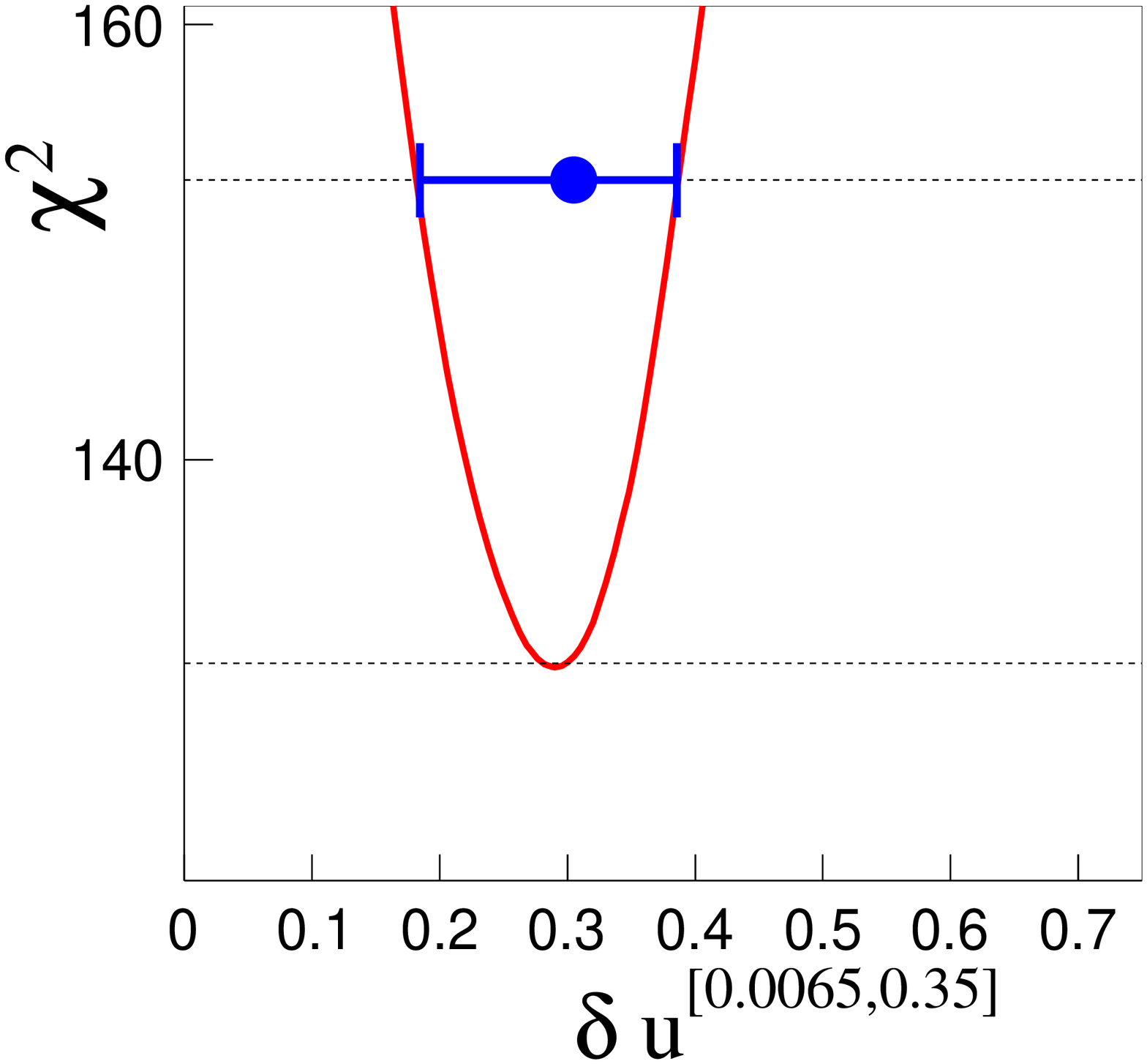}
\includegraphics[width=4.2cm,bb= 50 10 595 550]{\FigPath/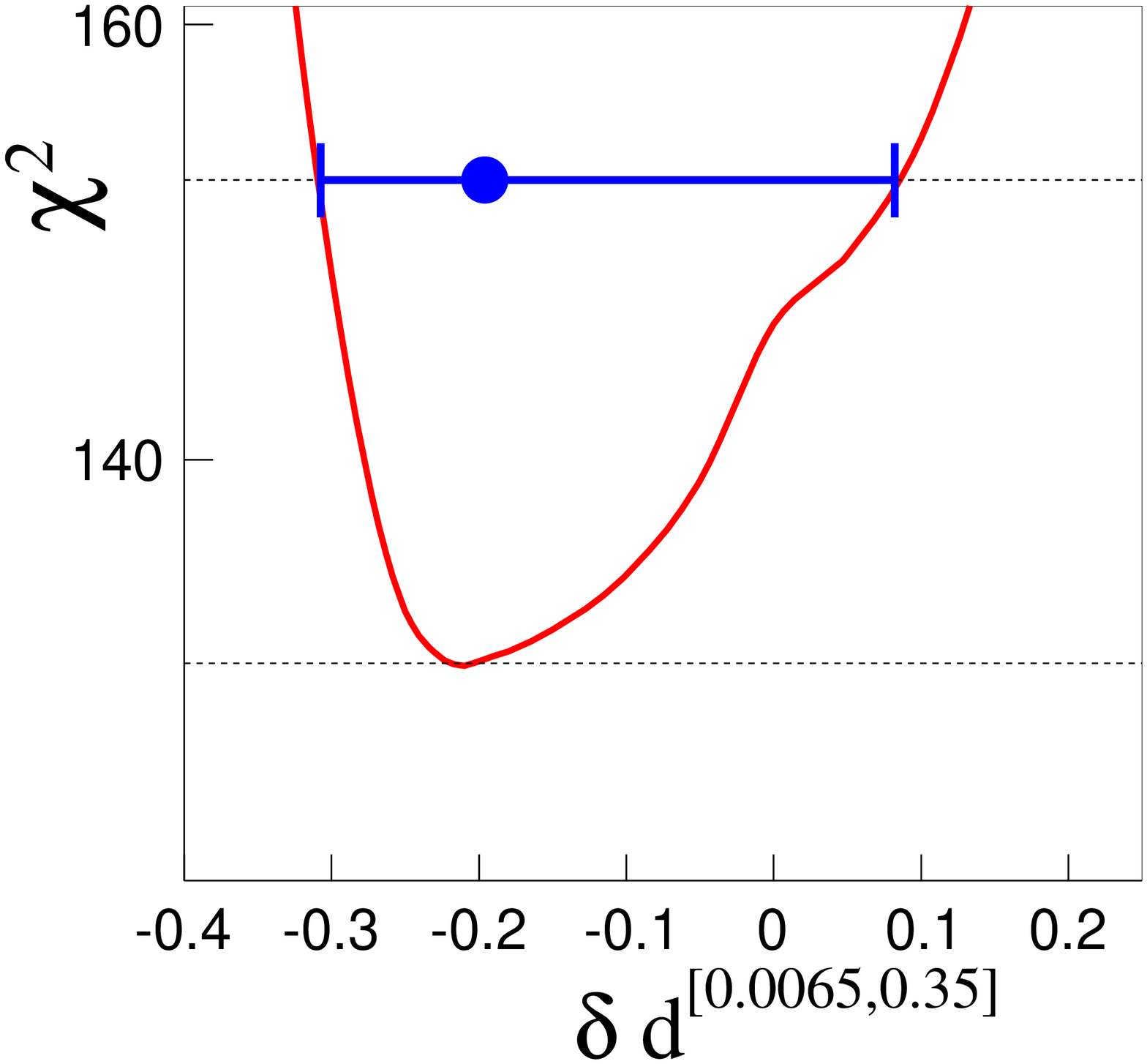}
\caption{$\chi^2$ profiles for up and down quark contributions to the tensor charge. The errors of points correspond to 90\% C.L. interval.}
\label{fig:chi2}
\end{figure}

Finally, we present an estimate at 90\% confidence level (C.L.) interval for  the
nucleon tensor charge contributions using the strategy outlined in
Refs.~\cite{deFlorian:2014yva,Martin:2009iq}. Transversity enters directly into SIDIS
asymmetry and we find that  the main constraints come from
SIDIS data only, its correlations with errors of Collins FF turn out
to be numerically negligible.
Since the experimental data has only probed the limited region $0.0065 < x_B < 0.35$,
we define the following partial contribution to the tensor charge
\begin{eqnarray}
\delta q^{[x_{\rm min},x_{\rm max}]}\left(Q^2\right) \equiv   \int_{x_{\rm min}}^{x_{\rm max}}dx \, h_1^q(x,Q^2) \ .
\end{eqnarray}
In Fig.~\ref{fig:chi2}, we plot the
$\chi^2$ Monte Carlo scanning of SIDIS data for the contribution to
the tensor charge from such a region, and find
\begin{eqnarray}
\delta u^{[0.0065,0.35]} &=&  +0.30^{+0.12}_{-0.08} \ ,\\
\delta d^{[0.0065,0.35]} &=&  -0.20_{-0.11}^{+0.28}   \ ,
\end{eqnarray}
at 90\% C.L. at  $Q^2=10$ GeV$^2$.
We notice that this result is comparable with previous TMD extractions
without evolution~\cite{Anselmino:2007fs,Anselmino:2008jk,Anselmino:2013vqa}
and the dihadron method~\cite{Bacchetta:2012ty,Bacchetta:2011ip}.

Existing experimental data covers a limited kinematic region,
thus a simple extension of our fitted parametrization to the whole range of $0<x_B<1$
will significantly underestimate the uncertainties, in particular, in the dominant large-$x_B$
regime.
It is extremely important to extend the experimental study of the quark transversity
distribution to both large and small $x_B$ to constrain the total tensor charge
contributions. This requires future experiments to provide measurements
at the Jefferson Lab 12 GeV upgrade~\cite{Dudek:2012vr} and the planned
Electron Ion Collider~\cite{Boer:2011fh,Accardi:2012qut}.

\section{Conclusions and outlook}
We have performed a global analysis of the Collins azimuthal asymmetries
in $e^+e^-$ annihilation and SIDIS processes, by taking into account the appropriate TMD
evolution effects at the NLL$'$ order and have constrained the nucleon tensor charge contributions from the valence up and down quarks in
the kinematics covered by the existing experiments .
The resulting transversity and Collins fragmentation functions
will be made available upon request in the form of a computer library. Future
developments will include  analysis of other spin asymmetries
including those from $pp$ scattering.
We emphasize the importance of future experiments
to further constraining the total tensor charge contribution of the nucleon.

\section{Acknowledgement}
We thank  D.~Boer, M.~Pennington, J.~Qiu, W. Vogelsang, and C.~-P.~Yuan for discussions and suggestions.
This material is based upon work supported by the U.S. Department of Energy,
Office of Science, Office of Nuclear Physics, under Contracts No.~DE-AC02-05CH11231 (P.S., F.Y.), No.~DE-AC52-06NA25396 (Z.K.), and No.~DE-AC05-06OR23177 (A.P.).


 \end{document}